\documentclass[11pt,nofootinbib]{revtex4}
\usepackage{amsmath,amssymb}
\usepackage[official]{eurosym}
\usepackage{setspace}
\usepackage[font={small,sf},format=plain,labelfont=bf,up]{caption}

\begin{document}
 
\title{Is it ``natural'' to expect Economics to become a part of the Natural 
Sciences?}

\author{Arnab Chatterjee}
\email[Email: ]{arnabchat@gmail.com}
\affiliation{Condensed Matter Physics Division, Saha Institute of Nuclear 
Physics, 1/AF Bidhannagar, Kolkata 700 064, INDIA}

\maketitle 

We are in the middle of a complex debate as to whether Economics is really a
natural science. The general consensus is that
\textit{natural science} concerns the description and understanding of 
natural phenomena, based on observations and empirical evidence. The two 
discrete branches are Physical and Biological sciences. Mathematics forms the 
basis of all of the above at different degrees of usefulness.
One very important feature of natural sciences is that the facts can be 
validated accurately, and the repeatability of the findings/results serve as  
fundamental criteria. Mathematics and logic form the basis of most natural 
sciences while observations and subsequently quantified facts become 
\textit{natural laws}. This is more prevalent in natural sciences than in 
social sciences. On the other hand, social sciences including economics are 
strongly based on qualitative reasoning, but at the same time, quantitative 
research forms a substantial part of it.

If we plunge deep into the philosophical aspects of science, the  
unique characteristic of a scientific methodology is the way of validating its 
ideas, commonly termed as `epistemology'~\cite{eichner1983economics}. 
The fundamental test of coherence for determining whether the derived 
conclusions follow from the assumptions logically and whether the arguments are 
consistent, is still considered by most theoretical economists to be 
sufficient. 
They favor logical analysis, and hence the exclusive use of mathematics, bound 
to ``theorems'' and ``proofs''. The test of coherence is a necessary but not a 
sufficient condition, when it comes to natural science, as established by 
scientists and philosophers. What are further required are the tests of 
\textit{correspondence}, \textit{comprehensiveness} and \textit{parsimony}, 
which are all empirical in nature. 
When all three of these tests are met, a theory can be said to have 
been validated empirically. 
Social scientists usually point out that it is very difficult to carry out 
empirical research and so, their theories should be accepted without
necessarily passing all these tests.
Specifically economists argue that it is difficult or rather impossible to 
conduct experiments where certain parameters can be controlled.
However, these issues asking to relax the mandatory empirical validation of a 
theory, is at conflict with the course of natural science.
It is until not quite recently that approaches have been changing slowly.

Being trained as a physicist, and having tried my hand on addressing some 
socio-economic problems, with my little experience and expertize, I will try to 
analyze the present question to the best of my ability. 
Here I will not make comments about political economy and related matters of policy making that I 
have very little idea about.
Moreover, in this article, I will argue about economics and social sciences in general.

In the last 2 decades, we have seen a surge in the number of physicists 
pursuing research in topics which traditionally belong to economics. So much so 
that even a term 
\textit{Econophysics}~\cite{stanley1996anomalous,Mantegna2000,
sinha2010econophysics} has been coined for the cross-disciplinary studies which 
use mainly physics approach and tools to address economics problems. This in 
turn triggered many meetings (conferences, symposia, workshops etc.) and 
subsequently academic journals to support this line of study. Several books 
have already been written by experts to formalize the course of these studies, 
and a few universities and institutions have been offering courses since 
more than a decade now, and even several Ph.D. theses have been written on the 
topics. This is quite encouraging for the interdisciplinary approach to science 
in general, 
although both the communities are still not quite sure of this `marriage' of 
physics and economics, since there is  a feeling that nothing 
substantial has been achieved compared to the amount of activity that has 
already taken place.
It is somewhat true that economists, being not so liberal to interdisciplinary 
approaches, do not pay much attention to whatever progress is offered by 
physicists to their subject.
On the other hand, we are often being asked by traditionally grounded 
physicists, whether \textit{Econophysics} is just a fad.
As the saying goes, ``\textit{the proof of the pudding is in the eating}'',
both communities will pay attention to the need for cross-disciplinary talking 
only when there is some serious contribution in decoding an ill-understood 
phenomena. From a lay physicist's point of view, several intuitively appealing 
concepts of economics such as efficiency and optimality are rather abstract, 
and difficult to formalize, and possible alternatives might help in the overall 
understanding of the subject.

Let us go back in time to see how the disciplines evolved.
Sociology and Economics are disciplines in their own right, with a huge body of 
modern literature developed independently of physical sciences. 
However, in their infancy, these disciplines were not very distinct
from physical science. It is interesting to note that the development of 
statistical physics was also 
influenced by social statisticians recording the `state' of a person 
(statistics as `state numbers') by 
recording the various measures of his social conditions.
Theoretical development of social science was triggered in the name of ``Social 
Physics'' \`{a} la Adolphe Quetelet~\cite{quetelet1835homme}, and subsequently 
by Auguste Comte~\cite{comte1856social}, 
The development of economic theory through the 19th century saw analogies to 
physics. One of the founders of modern economics, Irving Fisher was originally 
trained as a physicist under Josiah Willard Gibbs, who was in turn among the 
foremost to lay foundations of statistical mechanics.
Of course, one of the most well known facts 
in the history of modern 
economics is that many of its ideas have been largely influenced by physical 
sciences, with their logical basis 
and technicalities having close resemblance to statistical physics.
A classic example is that of Jan Tinbergen, who with his colleague Ragner 
Frisch, 
was the first Nobel laureate in Economics (Nobel Memorial Prize in Economic 
Sciences  in 1969)
``for having developed and applied dynamic models for the analysis of economic 
processes''.
Tinbergen studied mathematics and physics at the University of Leiden under Paul 
Ehrenfest,
who was one of the key figures in the development of statistical physics. 
During his years at Leiden, Tinbergen had numerous discussions with 
Kamerlingh Onnes, Hendrik Lorentz, Pieter Zeeman, and Albert Einstein,
all Nobel laureates who left profound contributions to statistical physics.
Tinbergen and many other icons of modern economics shaped their ideas
with heavy influences from physical science, most of which were already  
developed in the literature of statistical physics.
Another fact in the story of cross-influence is that of the history of 
the theory of random walk, which was formalized mathematically by Louis 
Bachelier, in his thesis \textit{Th\'{e}orie de la 
sp\'{e}culation}~\cite{bachelier1900theorie}, predating Albert
Einstein's~\cite{einstein1905} and Marian Soluchowski's~\cite{smoluchowski1906} 
study of Brownian motion.

Social scientists and economists have since then dealt with numerous
interesting issues of the human society, uncovering behaviors which seem
to be universal. 
These empirical regularities (or patterns) suggest that they might be 
predictable from first principles~\cite{farmer2005economics}, despite the fact 
that development of society and economics can be influenced heavily by 
historical events, that might seem to be an integral part to how the society 
functions.
Markets sit at the central point of the economy, where communications take 
place directly or indirectly, and there exists collective processes of price 
formation and resource allocation~\cite{chakraborti2015statistical}, and 
subsequent rise of social institutions to support all of these.
In the quest for the above, physicists and economists approach problems with 
different goals. As in most natural sciences, physics is propelled by the quest 
for ``universality''~\cite{binney1992theory}, what are usually termed as 
``laws''. The quest for similarity within a diverse set of events lead to this, 
where several details that make these sets seemingly different are 
eventually shown to be redundant or ``scalable'' in some appropriate 
mathematical form. Social science in general, as also in economics, is more 
likely to focus on the differences between the same sets.

However, it is also often pointed out that certain patterns 
observed a couple of decades ago may have changed. This happens due to the fact 
that certain essential features of human interaction have changed with the
advent of technologies. This makes socio-economic systems distinct 
compared to physical systems -- here we rarely find established 
`laws'~\cite{ball2004physical}.
This naturally calls for a change in the theoretical analysis
and modeling, making the field an extremely challenging one.
Specifically, there has been  wide discussion about addressing
socio-economic phenomena in the light of 
\textit{complexity science}~\footnote{M. Buchanan, To Understand Finance, 
Embrace Complexity, Bloomsberg, March 11, 2013. 
http://www.bloomberg.com/news/2013-03-10/to-understand-finance-embrace-complexit
y.html}
and embracing ideas from various 
disciplines~\cite{buchanan2013forecast}
to understand socio-economic phenomena
and a few big initiatives are already taking shape~\footnote{Institute for New 
Economic Thinking, http://ineteconomics.org/}.

Mandelbrot's finding of the power law behavior of price 
fluctuations~\cite{mandelbrot1997fractals} and further establishment of the 
results by different groups, including Gene Stanley's~\cite{gabaix2003theory} 
is one of quite a few important contributions made to financial markets.
Suddenly powers laws were found everywhere in the financial market data,
including the variance in the company growth rate with respect to company 
size~\cite{stanley1996scaling}, number of shares traded in a transaction, 
number of trading orders submitted at a specific price relative to the best 
offered price, size of the price response given the size of a company, 
among others. Pricing models have seen  sufficient 
contributions~\cite{bouchaud2000theory} from physicists, and studies of stock 
price movements in analogy to earthquake statistics~\cite{sornette2009stock}, 
have engaged the communities to debate.
Another important contribution came in the form of studying income and wealth 
distributions~\cite{Yakovenko:RMP,chakrabarti2013econophysics}. Since Pareto's 
observation~\cite{Pareto-book} of the power law in income distributions
back in 1897, most theories were confined to economic details. 
Physicists contributed  fresh eyes and a new toolbox, as well as 
the importance of several fundamental models like kinetic 
theory~\cite{Chatterjee2007} which sit at the base of physical sciences,  
to explaining the nature of income and wealth distributions.

Economists believed that the foundations of the subject was strong 
enough, yet the standard economic theory failed to foresee the graveness 
of financial crises that occurred in the last 10 years. Many started to express 
an opinion about the need for a fresh foundation that took into account the 
role of heterogeneous agents~\cite{lux2009economics}. Except a very few areas, there is not yet 
acceptance with open arms for such approaches from mainstream economics, where 
theories are mostly based on a representative agent, which are attributed to as 
the reason for the failure of such mainstream theories to predict the economic 
catastrophes.

More liberal approaches like agent based modeling, specifically focusing on 
heterogeneous agents, which almost intuitively corresponds to the formalisms 
developed for the theoretical physics of disordered 
systems~\cite{chaikinlubensky,sethna2006statistical}, specifically useful being 
the physics of spin glasses~\cite{binder1986spin}. Agent based modeling is 
dominated by game theory, of which, most of the theoretical studies are two-agent games,
while there is not sufficient literature on multi-agent, or for that matter, 
study of the case of macroscopically large number of agents. 
An alternative approach using  physics of 
disordered systems has proved to be useful, yet is not a typical economics 
tool. A culmination of several of these 
approaches~\cite{challet2004minority,chakraborti2015statistical} can offer 
useful alternatives to addressing a variety of problems.
Traditional economics is dominated by the assumptions of rational 
agents, and the `representative' one and their optimizing behavior, which 
are, time and again proved to be too ideal for the real world. 
A single agent, household or firm , maximizing their utility makes little sense.
Surely they make nice looking theories, with solvable equations, but are indeed 
far off from reality, as the empirical facts prove. Unless one accepts the 
reality of `interaction' between economic agents (interaction as in 
\textit{I react according to how you act}), one is not expected to 
understand the dynamics of collective activity of a large number of differently 
motivated, intelligent players with varied information processing 
capabilities who can, in principle, take a variety of actions.
There is already very strong arguments~\cite{lux2009economics} in favor of such 
an approach which might lead to rather useful things.

The answer to the basic question in our hand, whether \textit{Economics} is yet 
to qualify as a \textit{Natural Science} is still lurking, although both 
Economics and Natural Sciences have evolved a lot since such a question was 
formally raised~\cite{veblen1898economics} back in 1898, and repeatedly 
later~\cite{eichner1983economics,schabas2009natural} within the economics community.
The future is ours, it is never too late to start reshaping approaches for good.
Accepting ideas and methodologies from natural sciences into economics can only 
better the understanding and control over an economy.

\vskip 0.2cm \noindent 
Acknowledgement: I am grateful to Soumyajyoti Biswas, Anindya S Chakrabarti, Bikas K Chakrabarti, Asim Ghosh, 
Joanna Hutchinson, Sudip Mukherjee and Parongama Sen for useful comments, 
inputs and criticisms on this article.
%

\end{document}